\documentclass[12pt]{iopart}
\usepackage{graphicx}
\usepackage{dcolumn}
\usepackage{bm}
\usepackage{cite}
\usepackage{color}
\usepackage{xcolor}
\usepackage{comment}
\usepackage{mathtools}
\usepackage{iopams}

\begin{document}
\title[Unraveling Dynamic Patterns in Active Fluids with Nonlinear Growth]{From Order to Chimeras: Unraveling Dynamic Patterns in Active Fluids with Nonlinear Growth}
\author{Joydeep Das, Abhishek Chaudhuri$^{*}$ and Sudeshna Sinha}
\address{Department of Physical Sciences, Indian Institute of Science Education and Research Mohali, Sector 81, Knowledge City, S. A. S. Nagar, Manauli PO 140306, India}
\address{$^*$ Author to whom any correspondence should be addressed.}
\eads{\mailto{abhishek@iisermohali.ac.in}}

\begin{abstract}
  We explore pattern formation in an active fluid system involving two chemical species that regulate active stress: a fast-diffusing species ($A$) and a slow-diffusing species ($I$). The growth of species $A$ is modelled using a nonlinear logistic term. Through linear stability analysis, we derive phase diagrams illustrating the various dynamical regimes in parameter space. Our findings indicate that an increase in the P\'eclet number results in the destabilisation of the uniform steady state. In contrast, counter-intuitively, an increase in the nonlinear growth parameter of $A$ actually stabilises the homogeneous steady-state regime. Additionally, we observe that greater asymmetry between the species leads to three distinct dynamical phases, while low asymmetry fails to produce oscillatory instability. Numerical simulations conducted in instability regimes show patterns that range from irregular, arrhythmic configurations at high P\'eclet numbers to both transient and robust symmetry-breaking chimera states. Notably, these chimera patterns are more prevalent in the oscillatory instability regime, and our stability analysis indicates that this regime is the most extensive for high nonlinear growth parameters and moderately high P\'eclet numbers. Further, we also find soliton-like structures where aggregations of species $A$ merge, and new aggregations spontaneously emerge, and these patterns are prevalent in the phase of stationary instability. Overall, our study illustrates that a diverse array of patterns can emerge in active matter influenced by nonlinear growth in a chemical species, with chimeras being particularly dominant when the nonlinear growth parameter is elevated.
\end{abstract}

\maketitle

\section{Introduction}
Pattern formation is a ubiquitous phenomenon in natural systems. In particular, it is an integral part of the development of biological systems. The classical framework for modelling and understanding pattern formation is that of reaction-diffusion systems~\cite{turing1990chemical,cross2009pattern,cross1993pattern,kondo2010reaction,green2015positional}. The emergent patterns observed in this broad class of systems include time-independent and time-dependent oscillatory patterns~\cite{turing1990chemical}, symmetry-breaking instabilities~\cite{prigogine1968symmetry}, travelling waves~\cite{zaikin1970concentration}, spirals~\cite{winfree1972spiral} and jumping oscillations~\cite{knobloch2021origin}. Importantly, many of these patterns have been verified in experiments as well~\cite{ouyang1991transition,castets1990experimental}. 

However, in most biological systems, active transport and forces play a critical role in pattern formation. A prototypical approach to understanding pattern formation in biological systems, therefore, integrates the contributions of mechanical and chemical effects~\cite{harris1984generation}, for instance actin networks showing non-equilibrium dynamics by force generation through myosin motor activity~\cite{kumar2014pulsatory,juelicher2007active,ramaswamy2010mechanics} and pattern formation in active fluid medium in the presence of diffusing chemical species that are advected by self-generated flows produced by concentration-dependent active stress gradients \cite{bois2011pattern,kumar2014pulsatory}. Cell crawling is also dependent on the density of myosin motors, and so it can be studied by considering the distribution of myosin motors as a supercritical van der Waals (vdW) fluid~\cite{drozdowski2023optogenetic}. Spontaneous protrusion dynamics is modelled by mechano-chemical coupling via a polymerizing active gel layer~\cite{levernier2020spontaneous,laplaud2021pinching}.

The actomyosin cortex, which lies just beneath the cell membrane, consists of actin filaments and myosin motor proteins which crosslink the actin. This generates mechanical forces, giving rise to an active stress component in a thin mesoscopic layer of the actomyosin cortex. The actomyosin cortex is treated as an active fluid at time scales of morphogenesis, which is an essential part of organismal development, relevant to human tissue development as well as development in systems such as worms, flies, zebrafish and mice~\cite{kruse2024actomyosin}. The active stress can now be considered to be a function of the concentrations of regulatory chemical species to complete the mechanochemical integration, leading to pattern formation even in the absence of chemical reactions~\cite{bois2011pattern}. When extended to two chemical species which undergo advection-diffusion and regulate the active stress, this framework gives rise to pulsatory patterns~\cite{kumar2014pulsatory}. Spontaneously emerging localized states in the active fluid medium have led to the understanding of localized cellular patterns~\cite{barberi2023localized}. It has been shown that localized states can emerge spontaneously (analogous to isolated clusters of actin and signalling molecules in cancer cells~\cite{dd2061aa35da449e91a2a6209d0118ab}) if the assembly of active matter is regulated by the presence of chemical species that are advected with flows resulting from active stress gradients. 

In another research direction, in the context of dynamical systems in general, a class of patterns that have generated intense research interest are chimera states. The scope of the term ``chimera'' has vastly expanded since it was first introduced \cite{chimera}, and a number of variations such as frequency chimeras, amplitude and amplitude mediated chimeras, breathing chimeras, traveling phase-cluster chimeras and multiple phase-cluster chimeras, have been characterized. These states have had alternate descriptors in earlier papers such as clustered states, or coexisting states. Broadly, a chimera state refers to one where a system spontaneously breaks the underlying symmetry and splits into co-existing groups that have significantly different dynamical features \cite{chimera1,chimera_ss,punit}, i.e. the emergent pattern has spatial domains exhibiting distinct temporal behaviour. The term ``chimera-like'' has also been widely used to refer to non-stationary chimera states with irregular phase boundaries \cite{chimera_like_1,chimera_like_2}. This fascinating phenomenon has been observed in a variety of systems, ranging from Josephson junction arrays \cite{chimera16} to uni-hemispheric sleep in certain animals \cite{chimera18}, as well as in continuous media models \cite{ALVAREZSOCORRO2021105559, PhysRevLett.119.244101}. Importantly, these chimera states have also been observed experimentally in optical analogs of coupled map lattices  \cite{chimera9}, Belousov-Zhabotinsky chemical oscillator systems \cite{chimera8}, two populations of mechanical metronomes \cite{chimera19}, electronic circuit systems \cite{star, chimera20}, neurodynamics \cite{compte2000synaptic} and Liquid Crystal Light Valve experiments with optical feedback \cite{verschueren2013spatiotemporal}. So, it is of immense interest to uncover chimera patterns in different classes of systems to gauge the generality of this interesting spatiotemporal pattern. 

Motivated by these broad ideas, in this paper, we show the emergence of chimera states in an active fluid with two chemical species regulating the active stress as in Ref.~\cite{kumar2014pulsatory}, with one species showing a logistic growth. Our work, while generic for any two-component advection-diffusion system coupled to an active fluid, is more specifically related to the actomyosin cortex. Pulsatile patterns have been observed in several situations in the actomyosin cytoskeleton. In that context, myosin motor proteins could represent the species that up-regulates stress. Any of the several proteins associated with the actomyosin cortex - like actin-monomer binding proteins, severing proteins, cross-linking proteins or filament binding proteins - could serve as the species which down-regulates the stress. The logistic growth term also has a simple explanation. In pattern formation in actin networks, the growth term arises naturally to describe the growth of existing F-actin fibres and nucleation of new fibres, while crowding will slow down polymerization to justify the satuaration~\cite{le2016pattern}. Separately, a generic mechanism involving actin turnover and myosin activity in the ring formation in the Drosophila trachea also necessitates using a variant of the non-linear growth term~\cite{hannezo2015cortical}. Both these settings fall within the necessary requirements in our model.

A logistic growth term has also been used extensively in many mathematical models of biological systems and provides a general description of population growth. In particular, a logistic growth model with long-range interactions serves as a generic minimal model for competition for common resources and pattern formation in excitable media~\cite{shnerb2004pattern}. For instance, in the chemotaxis model proposed in Ref.~\cite{painter2011spatio}, the logistic term is used to model the growth of cell density. In autochemotactic pattern formation, logistic growth models the death and reproduction of self-propelled bacteria ~\cite{mukherjee2018growth}. Further, the logistic functional form has been used in an experimentally motivated model of the coupling between reaction-diffusion and active matter~\cite{PhysRevE.105.014602}, as well as in gene expression dynamics~\cite{fisher1937wave,neufeld2009chemical}. Polymerization in actin gels
and solutions is also expressed by a reactive logistic term, useful to describe many characteristic states of actin-wave formation: spots, spirals, and travelling waves~\cite{yang2014spiral,le2016pattern}.

Our central questions in this work are two-fold: First, is nonlinearity detrimental to the stability of homogeneous steady-state states in these active fluid systems, or does the nonlinear logistic growth of a chemical species aid the stability of such regular states? Secondly, we seek to determine the classes of spatiotemporal patterns that emerge in the system when the uniform steady-states lose stability. Specifically, we will search for interesting patterns, such as chimeras, in the space of varying parameters, including the strength of the logistic growth term. That is, we will seek to identify the generic conditions that underlie the transient or robust generation of chimera-like patterns in active fluids. This study then will potentially lead to a better understanding of the effects of the nonlinear growth of a chemical species on emergent dynamical patterns in the active fluid medium and has the scope to motivate engineered in-vitro experiments of reaction-diffusion and active matter systems~\cite{PhysRevE.105.014602,padirac2013spatial,chirieleison2013pattern,vyborna2021dna,senoussi2021programmed,nguindjel2022spatial,zaferani2024building,zadorin2015synthesis,rajasekaran2024programmable}. 
  
In the Sections below, we will first describe the model and then present the stability analysis of the system. We will then explore the emergent patterns through numerical simulations over a wide range of parameters. In particular, we will demonstrate the existence of chimera states, solitonic defects and merging-emerging patterns.

\section{Mathematical Model}

Consider two distinct chemical species ($A, I$) regulating the active stress in a one-dimensional active fluid medium in a thin-film geometry. The quantities of interest are the concentration fields of $A(x,t)$ and $I(x,t)$ at time $t$ and position $x$. The evolution equations of the two chemical species in one-dimension are given as:
\begin{eqnarray}
\partial_t A &=& - \partial_x (vA) + D\partial^2_x A + rA(1 - A/K)\\ 
\partial_t I &=& - \partial_x (vI) + \alpha D\partial^2_x I
\end{eqnarray}
Thus, both species undergo advection and diffusion, with the diffusive component determined by the diffusion coefficient $D$ and the advective component given by the bulk fluid flow velocity $v$. The relative diffusion coefficient of the two species is determined by the parameter $\alpha > 0$.
The nonlinear logistic growth term, with strength $r>0$, has the capacity to destabilize a steady state with zero concentration and also yields saturation at a finite carrying capacity $K$. For the active fluid at low Reynolds numbers, the inertial terms can be neglected, and the force balance equation gives
\begin{equation}
\partial_x\sigma = \gamma v,
\end{equation}
with
\begin{equation}
\sigma = \eta\partial_x v + \sigma_a
\end{equation}
\\
giving the total stress. $\sigma$ consists of a viscous stress component $\eta\partial_x v$ with $\eta$ being the viscosity of the medium and an active stress component $\sigma_a$ regulated by the concentrations of the chemical species:
\begin{equation}
\sigma_a = \sigma_0 f(A,I)
\end{equation}
where $\sigma_0$ is an active stress amplitude. $f(A, I)$ is a dimensionless function describing the regulations of the active stress:
\begin{equation}
f(A,I)= (1+\beta)\frac{A}{A+A_S}+(1-\beta)\frac{I}{I+I_S}
\label{f_A_I}
\end{equation}
where $\beta$ is an asymmetry parameter, and $A_S$ and $I_S$ represent the saturation values of the concentrations of the two chemical species~\cite{kumar2014pulsatory}. For $\beta < -1$, $A$ down-regulate and $I$ up-regulates active stress; for $-1\leq \beta \leq 1$ both species up-regulates stress and for $\beta > 1$, $A$ up-regulates while $I$ down-regulates active stress. 
For $A < A_S$ and $I < I_S$, this functional form is linear in leading order, and a simple linear functional form of $f(A, I)$ also exhibits qualitatively similar spatiotemporal patterns.

For further analysis, Eqs.(1)-(4) are expressed in dimensionless form:
\begin{eqnarray}
    \partial_t A = - \partial_x (vA) + \partial^2_x A + RA(1 - A)\\ 
\partial_t I = - \partial_x (vI) + \alpha \partial^2_x I\\
\partial^2_x v + Pe\partial_x f(A,I) = v
\end{eqnarray}
where $A, I, x, t$ and $v$ are now nondimensional and the scaled non-linear growth parameter $R = r\eta/\gamma D$ and the dimensionless P\'eclet number $Pe = \sigma_0/\gamma D$.

\section{Stability Analysis}

\begin{figure*}
\centering
\includegraphics[width=0.85\textwidth]{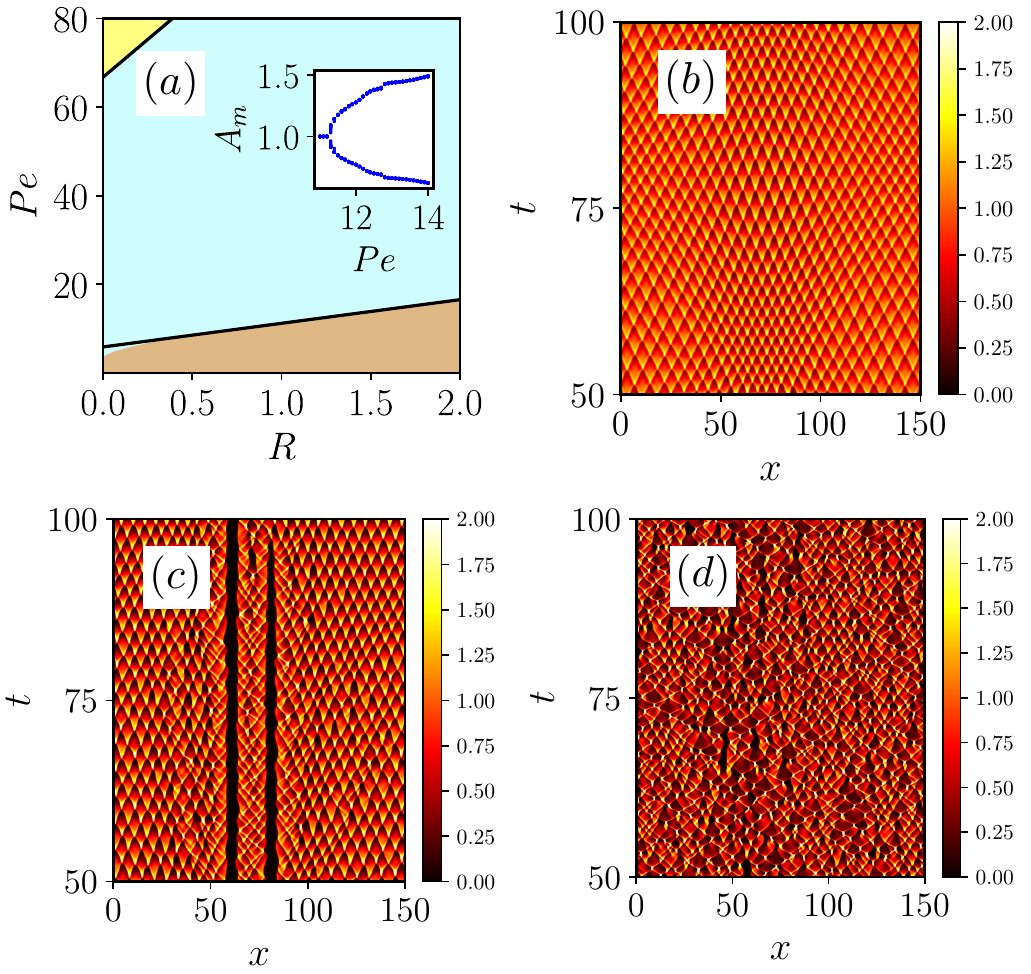}
\caption{(a) Phase diagram of the different dynamical patterns that emerge, in the parameter space of the P\'eclet number $Pe$ and the scaled nonlinear growth parameter $R$, obtained from Linear Stability Analysis. Here the ratio of diffusion constants $\alpha=0.1$ and the asymmetry parameter $\beta=3$. The brown colour represents the stable homogeneous steady state, pale blue represents oscillatory instability, and yellow represents stationary instability. The black curve represents the linear boundary between the stable homogeneous steady state and the regime of oscillatory instability, given by Eqn.~\ref{linear_approx}. Further, the best-fit linear boundary, $Pe \sim 34 R$, between the regime of oscillatory instability and stationary instability is also demarcated in black. (Inset) Numerically obtained bifurcation diagram, with $Pe$ varying from $11$ to $14$ ($\beta$=$3$, $\alpha$=$0.1$, $R=1$), clearly showing the transition from homogeneous steady states to oscillatory regime, through a supercritical Hopf bifurcation. Here, $A_m$ represents the local extrema of $A$ at a particular site after a long transience. (b-d) Kymographs obtained from numerical simulations, illustrating the emergent spatiotemporal patterns in the region of instability, for $R=1$, as the P\'eclet number increases from left to right: (b) $Pe=18$ (regular pulsatory state), (c) $Pe=35$ (chimera state) and (d) $Pe =48$ (spatiotemporal chaos). In the kymographs, the concentration $A$ is displayed as a function of time (vertical axis) and space (horizontal axis), with the colour bar indicating increasing concentrations from $A = 0$ (black) to $A=2$ (white).
    }
    \label{Pe_R}
\end{figure*}

Consider homogeneous steady states, with concentrations $(A_0,I_0)$, and velocity $v=0$. We consider perturbations to the homogeneous steady state $(A_0, I_0)$ of the form $A=A_0+A_p(x,t)$, $I=I_0+I_p(x,t)$ and $v=v_0+v_p(x,t)$, where, $A_p(x,t)=\delta A_0 e^{\lambda t + i k x}$, $I_p(x,t)=\delta I_0 e^{\lambda t + i k x}$, $v_p(x,t)=\delta v_0 e^{\lambda t + i k x}$. The important parameters here are (1) the P\'eclet number, which gives the ratio of the diffusive time scale and the advective time scale, (2) the parameter $R$, which reflects the strength of nonlinearity, and (3) $\beta$, which reflects the asymmetry in the two chemical species and influences evolution through the function $f$. Linear stability analysis of Eqs.~(7)-(9), yields the following Jacobian:
\begin{eqnarray}
\nonumber
J &=&-k^2
\begin{pmatrix}
  1 & 0 \\
  0 & \alpha \\
\end{pmatrix}
+ R(1-2A_0)
\begin{pmatrix}
  1 & 0 \\
  0 & 0 \\
\end{pmatrix} \\
&+& \frac{Pe k^2}{(1+k^2)}
\begin{pmatrix}
  A_0f_A & A_0f_I \\
  I_0f_A & I_0f_I \\
\end{pmatrix}
\end{eqnarray}
where $f_A, f_I$ are the partial derivatives of $f$ with respect to $A$ and $I$ respectively, evaluated at the homogeneous steady state. In what follows we consider $(A_S, I_S) = (3A_0, 3I_0)$ in accordance with the choice in Ref.~\cite{kumar2014pulsatory}. Note that the qualitative nature of the emergent patterns is not sensitive to this choice.

For each value of wave number $k=\frac{2\pi n}{L}$, $n=0, 1, \dots$ (considering periodic boundary conditions), we obtain two eigenvalues, which determine the stability and nature of the emergent dynamical pattern. The trace and the determinant of the Jacobian are given as:
\begin{eqnarray}
    Tr(J) &=&-k^2(1+\alpha) -R + \frac{Pe k^2 (f_A+f_I)}{(1+k^2)} \\ \nonumber
    \Delta(J) &=& k^4\left[\alpha-\frac{Pe(\alpha f_A+f_I)}{(1+k^2)}\right]\\ &&+Rk^2\left[\alpha-\frac{Pe f_I}{(1+k^2)}\right].
\end{eqnarray}
For the functional form given in Eq.~\ref{f_A_I}, evaluating the partial derivatives at the steady state, we obtain $f_A = (1+\beta) \frac{A_S}{(A+A_S)^2} = \frac{3}{16} (1+\beta)$ and $f_I = (1-\beta) \frac{I_S}{(I+I_S)^2} = \frac{3}{16} (1- \beta)$. So $f_A$ is always positive, while $f_I$ is negative for $\beta > 1$, and positive for $\beta <1$. Further notice that for $k \sim 0$, $Tr(J) \sim -R$, and for large $k$ it is negative as well. However, for intermediate $k$ it may be positive for certain parameter values, with the maximum value of $Tr(J)$ occurring at $k_{max}^2 = \sqrt{\frac{ 3 Pe}{8(1+\alpha)}} - 1$ (for $Pe > 8(1+\alpha)/3)$. For instance, for $\alpha=0.1$, this implies that the maximum of the sum of the eigenvalues occurs around $k_{max} \sim 1$ for $Pe \sim 11.7$.

Now, for a homogeneous steady state to be stable under perturbation, the disturbances from the homogeneous profile must damp exponentially rapidly in time. So the real parts of the eigenvalues for all $k$ should be less than zero, implying that $Tr(J) < 0$ and $\Delta(J) > 0$. 
In the parameter regime where the uniform steady state is unstable, the real part of at least one of the eigenvalues is positive, leading to the formation of diverse patterns. We denote the regime where the leading eigenvalues may not be imaginary as a stationary instability, in analogy with the terminology used in the Turing model. If the eigenvalues are imaginary for some values of $k$, an {\em oscillatory instability} will emerge.

\begin{figure*}[htb]
\centering
\includegraphics[width=0.85\textwidth]{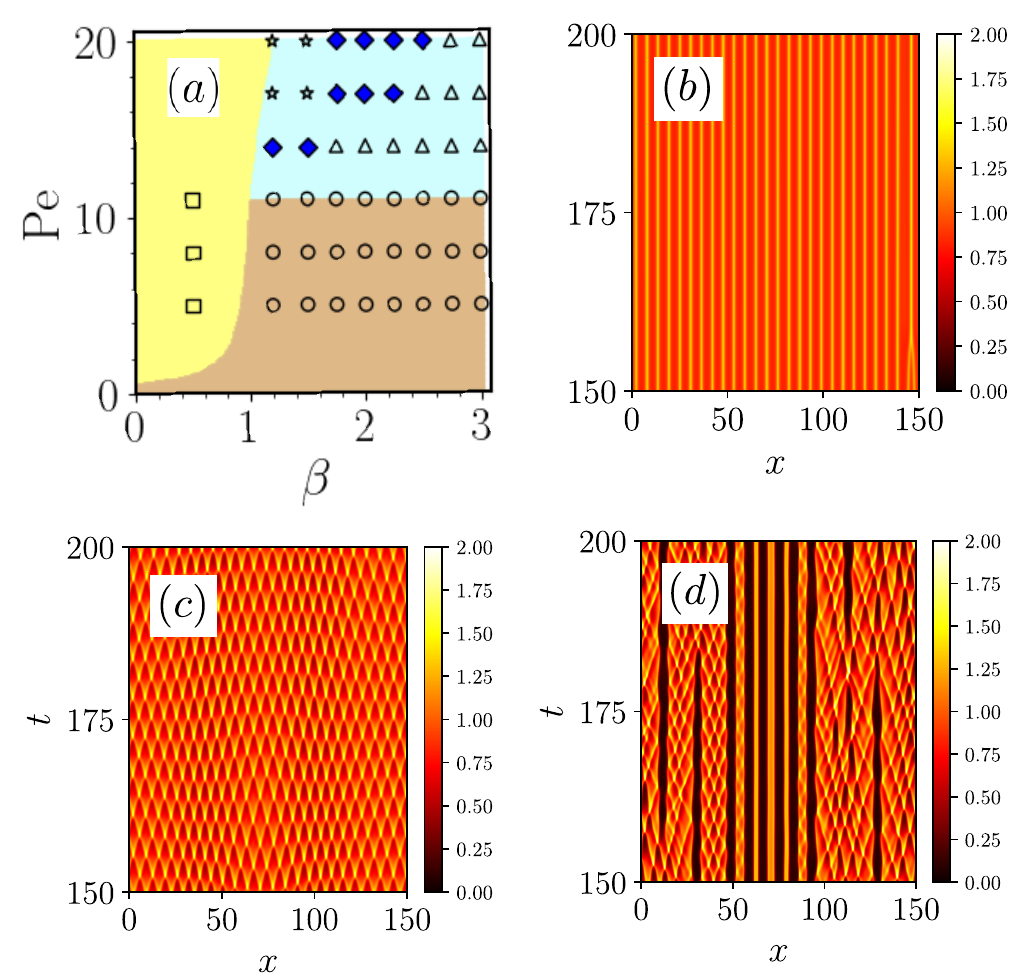}
  \caption{(a) Phase diagram of the different dynamical patterns that emerge, in the parameter space of the P\'eclet number $Pe$ and asymmetry parameter $\beta$, obtained from Linear Stability Analysis, for $\alpha$=$0.1$ and scaled nonlinear growth parameter $R=1$. The brown colour represents the stable homogeneous steady state, pale blue represents oscillatory instability, and yellow represents stationary instability. Symbols in the phase diagram denote the dynamical patterns obtained through numerical simulations: circle represents homogeneous steady states, square represents stationary Turing-like patterns,  triangle represents pulsatory patterns, {diamond represents chimera states and star represents merging-emerging dynamics}. (b-d) Kymographs obtained from numerical simulations illustrating the emergent spatiotemporal patterns in the region of instability: (b) $\beta=0.5, Pe=5$ (stationary patterns, denoted by squares in the phase diagram); (c) $\beta=1.75$, $Pe=14$ (pulsatory pattern, denoted by triangles in the phase diagram); (d) $\beta=1.75$, $Pe=17$.
    }
    \label{Pe_beta}
\end{figure*}

\begin{figure}
\centering
\includegraphics[width=0.85\textwidth]{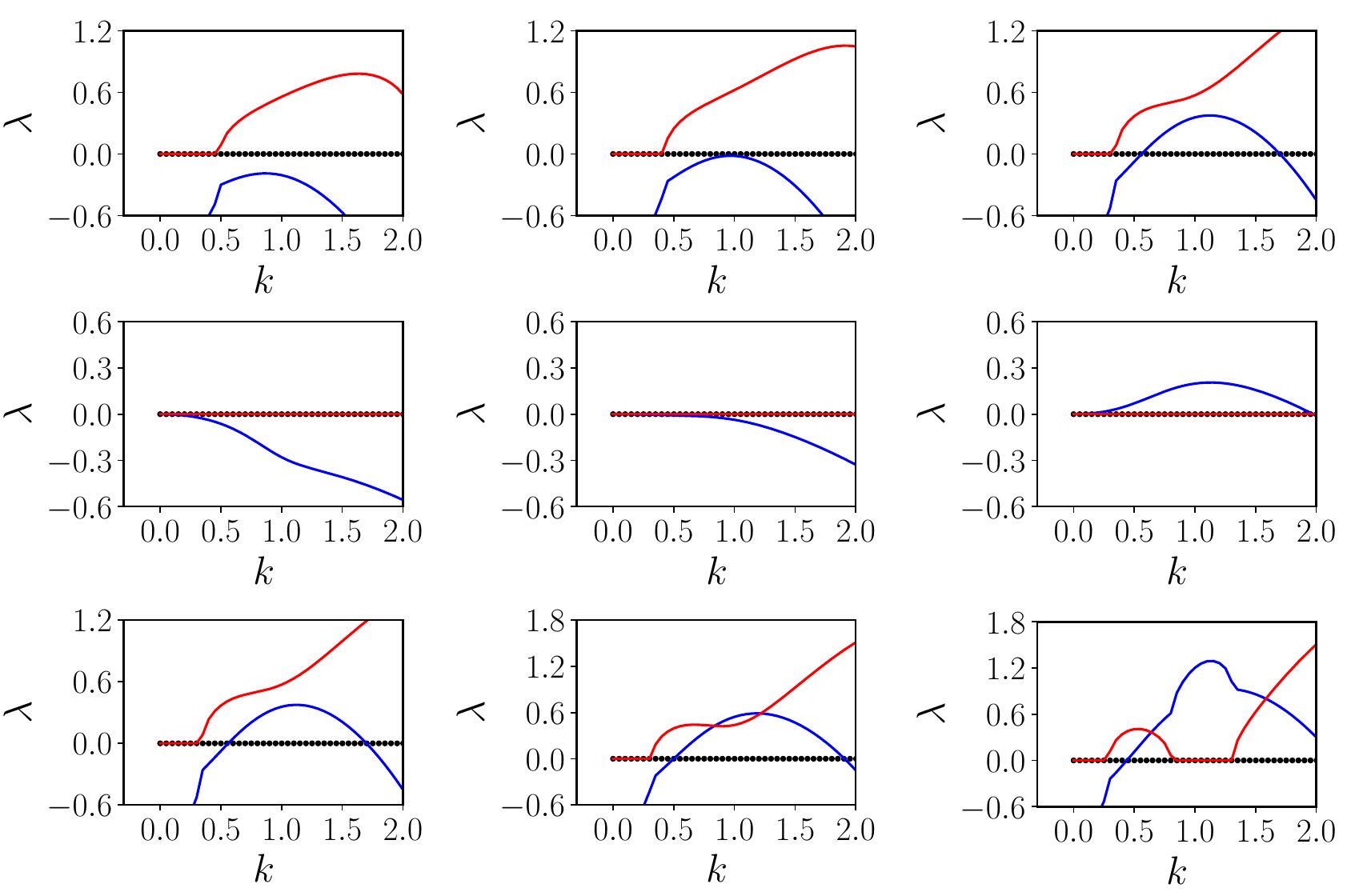}
    \caption{Dispersion curves obtained from Linear Stability Analysis, for varying P\'eclet number $Pe$ and  asymmetry parameter $\beta$, illustrating the transitions between the different dynamical regimes. Here $\alpha=0.1$ and nonlinearity parameter $R=1$. The top row of panels show the transition from the stable homogeneous steady state to the oscillatory instability, as $Pe$ increases from $9$ to $15$, for $\beta =1.2$. The middle row of panels shows the transition from the stable homogeneous steady state to stationary instability as $\beta$ decreases from $1.1$ to $0.7$ for fixed P\'eclet number $Pe=6$. The bottom row of panels shows the transition from oscillatory instability to stationary instability, as $Pe$ increases from $13$ to $20$, for $\beta =1.2$. In all panels, the blue curve represents the real part of the leading eigenvalue, the red curve represents the imaginary part of the leading eigenvalue, and the black dotted line shows $\lambda=0$ for reference.
    }
    \label{dispersion}
\end{figure}

We generate phase diagrams in the space of different parameter sets (cf Figs~\ref{Pe_R}-\ref{Pe_beta}) considering the variation of the eigenvalues with respect to wavenumber $k$, and noting the sign of the real part of the leading eigenvalue (see dispersion curves in Fig.~\ref{dispersion} which illustrate the behaviour of the real and imaginary parts of the leading eigenvalue when a parameter crosses a critical value giving rise to different classes of transitions). The three distinct dynamical phases, namely the stable homogeneous steady state, the regime of oscillatory instability and the regime of stationary instability, are marked with different colors in the phase diagrams. 

In Fig.~\ref{Pe_R}, we display the phase diagram in the parameter space of the P\'eclet number $Pe$ and the scaled nonlinear growth parameter $R$.  It is apparent from the phase diagram that as the P\'eclet number increases beyond a critical value, the homogeneous steady state gets destabilized, giving way to oscillatory instability. Inspecting the behaviour of the leading eigenvalue as the P\'eclet number crosses the critical value reveals a supercritical transition from the stable homogeneous steady state to oscillatory instability. On further increase of the P\'eclet number, the oscillatory instability changes to a stationary instability. 

In contrast, interestingly, the phase diagram also shows that the parameter space occupied by the homogeneous steady state is enlarged under increasing $R$, i.e. larger nonlinear growth yields enhanced homogeneity. This indicates a counter-intuitive stabilizing effect of increasing non-linearity on the dynamics. Further, increasing the magnitude of the nonlinear growth term also increases the parameter region supporting oscillatory states. It is also distinctly evident from Figure~\ref{Pe_R} that the boundaries between the different dynamical regimes have an approximate linear dependence on the scaled nonlinearity $R$. The slope of the linear rise of the boundary curve separating the homogeneous steady state and the oscillatory state is gentler than that of the boundary between the stationary instability and oscillatory instability. This implies that the oscillatory state increases most significantly with increasing nonlinear growth.

Specifically, our analysis also yields the boundary curve between the stable homogeneous steady state and the oscillatory instability. Along this boundary curve, the condition $Tr(J)=0$ should be satisfied.  So from the expression in Eqn.~11, we have $Tr(J) =-k^2(1+\alpha) -R + \frac{Pe k^2 (f_A+f_I)}{(1+k^2)}=0$, i.e. $k^2(1+\alpha) + R = \frac{Pe k^2 (f_A+f_I)}{(1+k^2)}$.
Now, as in the phase diagram in Fig.~1, we consider the values of $\alpha=0.1$ and $\beta=3$ (which gives $f_A+f_I=\frac{3}{8}$). Further, taking $k=1$ (since $k_{max} \sim 1$ over a large range of P\'eclet numbers) we obtain: $(1+\alpha) + R = \frac{3 Pe}{16}$. This yields the functional relation between the critical $Pe$ number and $R$ to be:
\begin{equation}
Pe_c=\frac{16}{3}(1.1+R)
\label{linear_approx}
\end{equation}
This linear relation is indicated by a line in the phase diagram displayed in Fig.~\ref{Pe_R}, and it is clear that this approximate expression fits results from stability analysis over a large range of $R$. Since the critical value of the P\'eclet number increases almost linearly with the scaled nonlinear growth parameter $R$, it implies that the range of $Pe$ over which the homogeneous steady state is stable is directly proportional to $R$, with larger stable ranges emerging for larger $R$.

When the system is weakly nonlinear, the critical value of the $Pe$ number also becomes very small, with  $Pe_c = 5.87$ for $R \to 0$, from Eqn. 13. However, notice that there is a small deviation from the linear relation given in Eqn. 13, in the phase boundary in Fig. 1, in the limit of weak nonlinearity. This can be rationalized as follows: Eqn. 13 was obtained using the approximation that $k_{max} \sim 1$. However, for very low $Pe$, this approximation is not accurate. Using the exact explicit expression for $k_{max}$ in place of the approximation yields a nonlinear equation for the critical $Pe$ number. Solving this nonlinear equation numerically (or graphically) gives a more accurate estimate of the critical $Pe$ number for weak nonlinearity, with $Pe_c = 8(1+\alpha)/3 = 8.8/3 \sim 3$ for $R \to 0$. This is completely consistent with the phase diagram in Fig. 1. For higher nonlinearity $R$, $Pe_c$ is also larger, and the linear relation in Eqn. 13 then offers an excellent approximation to the behaviour of the stability boundary.
 
We also observe that the pattern formation is crucially dependent on the asymmetry parameter $\beta$. This is clearly evident from Figure~\ref{Pe_beta}, where we display phase diagrams in the parameter space of the P\'eclet number $Pe$ and asymmetry $\beta$ for different values of the scaled logistic growth parameter $R$. When $\beta$ is low ($\beta < 1$), no regime of oscillatory instability exists for any value of the P\'eclet number, while for $\beta$ greater than $1$, all three dynamical regimes are possible. Another interesting observation is the independence of the phase boundary on the asymmetry parameter $\beta$, after $\beta \sim 1$ in Fig.~\ref{Pe_beta}. This can be rationalized as follows: the parameter $\beta$ enters the expression for the Jacobian through $f_A$ and $f_I$. Now the Trace is actually independent of $\beta$, as $(f_A+f_I)= 3/8$, making the expression in Eq.~11 independent of $\beta$. The Determinant, however, involves $\beta$. When $\beta < 1$, $f_I=\frac{3}{16} (1-\beta)$ is positive, and so the stability boundary is dependent on $\beta$ through the sign of the Determinant which is determined by the relative magnitudes of the positive and negative terms in Eqn.~12. On the other hand, for  $\beta>1$, $f_I$ is always negative, and so the sign of the Determinant is always positive after $\beta \sim 1$, in the limit of small $\alpha$. This implies that after $\beta \sim 1$, $\beta$ no longer influences the eigenvalues, and so the critical $Pe$ number is independent of it.

Further, sets of dispersion curves are also displayed in Fig.~\ref{dispersion}, clearly showing the behaviour of the real and imaginary parts of the leading eigenvalues when $Pe$ and $\beta$ cross critical values, underlying different classes of dynamical transitions. In particular, these dispersion curves illustrate the transitions from the stable homogeneous steady state to oscillatory instability on increasing $Pe$, and to the stationary instability on decreasing $\beta$. Note that the dependence of the dynamics on the asymmetry $\beta$ can be rationalized as follows: a negative determinant indicates that the signs of the two eigenvalues are different, and so a negative $\Delta(J)$ cannot yield imaginary eigenvalues (as they necessarily have to be complex conjugates as the Jacobian matrix is real). This implies the result that for small $\alpha$, $\beta > 1$ can yield $\Delta (J) > 0$, and so there can be modes that have oscillatory instability only for $\beta > 1$.

To summarize our results from stability analysis: first, increasing P\'eclet number leads to the loss of stability of the uniform steady state. Increasing asymmetry $\beta$ yields three distinct dynamical phases, while low asymmetry does not yield oscillatory instability. Lastly, interestingly, {\em nonlinear logistic growth aids the stability of the steady state and also enlarges the parameter regime supporting the phase with oscillatory instability.}

\section{Numerical Simulations}

We now explore the spatio-temporal dynamics of the system through extensive numerical simulations over a wide range of parameters. Note that a computational approach is necessary to examine patterns beyond steady states and simple pulsatory or stationary patterns. For instance, spatio-temporal defects and chimera states cannot be gauged by linear stability analysis alone. Further, numerical simulations provide valuable verification and consistency checks with analytical results, and they complement the analytical treatment of the system given above.

We solve Eqns.~(7)-(9) numerically by using a semi-implicit spectral scheme using periodic boundary conditions. We apply a small localized perturbation to the uniform steady state $(A_0, I_0)=(1,1)$, with $v=0$, and follow the system's response to this perturbation. Such localized perturbations help us track the spread of a small disturbance in the system and provide a method to gauge the response of the spatially extended system. With no loss of generality with respect to the qualitative features, we consider $L=150$, $\Delta x=L/512$, $\Delta t=0.0001$. Additionally, we have also examined the dynamics with $\Delta x=L/1024, L/2048$ in order to check the robustness of our qualitative results. We find that the emergent patterns are qualitatively robust with respect to system size. Further note that spatiotemporal patterns qualitatively similar to those displayed in our figures emerge from a large sample of random initial conditions at the specified parameter values.

\begin{figure*}
\includegraphics[width=\textwidth]{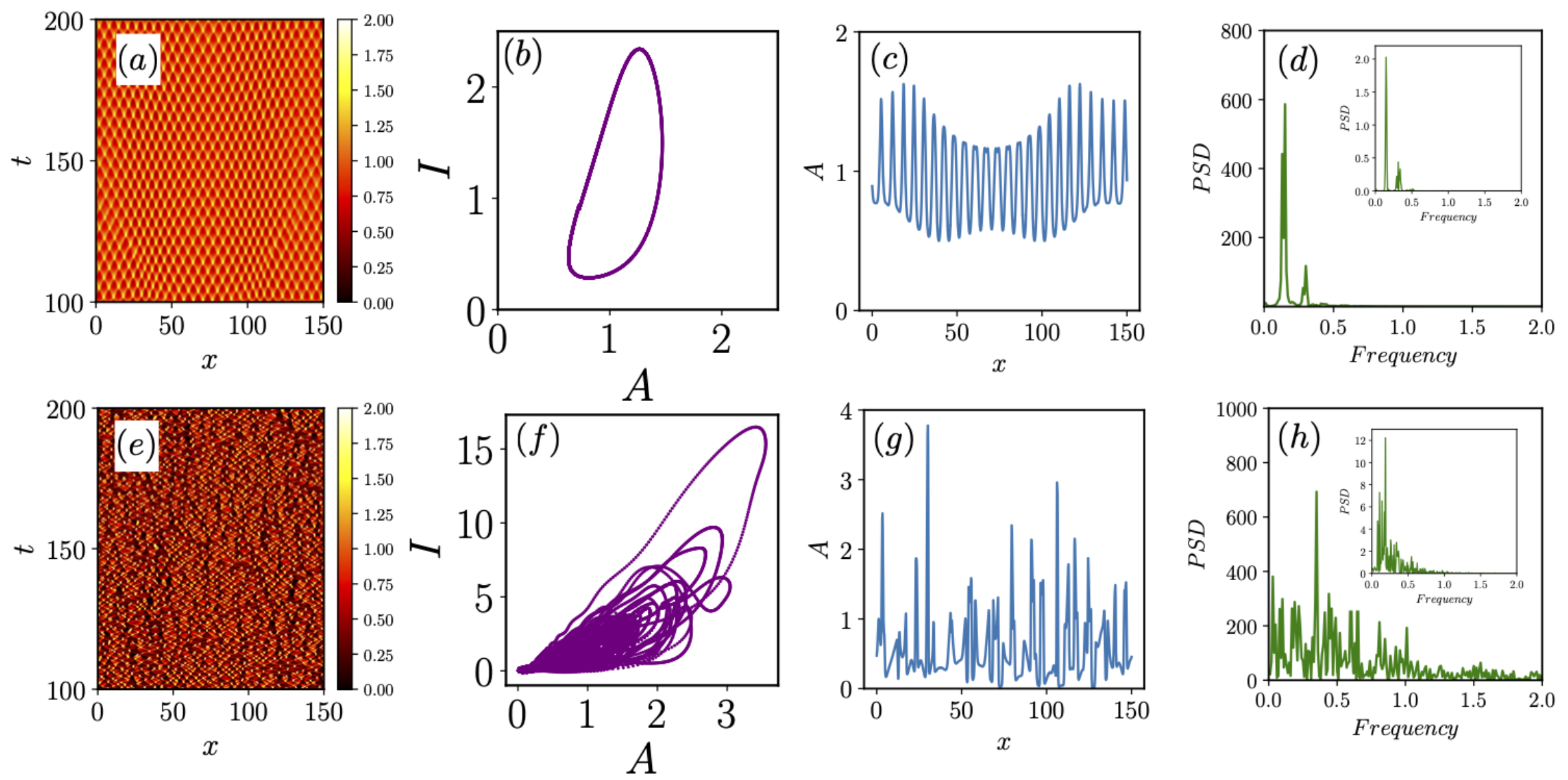}
\caption{[(a),(e)] Kymographs displaying the spatiotemporal evolution of concentration $A$; [(b),(f)] dynamical attractors in the phase space of $A-I$ concentrations at a typical location in space; [(c),(g)] snapshot of the spatial profile; and [(d),(h)] power spectrum (PSD) of the time series of $A$ at a site with spatial PSDs shown in the insets. Upper row corresponds to the parameters $Pe=13$, $\beta=1.5$, $R=1$, $\alpha=0.1$, and lower row corresponds to the $Pe=48$, $\beta=3$, $R=1$, $\alpha=0.1$. (b) and (c) display regular limit cycle behaviour and a regular oscillatory spatial profile. (f) and (g) display a chaotic attractor corresponding to the irregular spatial profile. Note that the system evolves to such attractors from a wide range of initial states, after transience.}
        \label{attractors}
\end{figure*}

Now we will demonstrate the wide variety of spatiotemporal patterns that arise from varying strength of the logistic growth term $R$, asymmetry $\beta$ and the P\'eclet number $Pe$. 
The first key observation is that the homogeneous steady state loses stability as the P\'eclet number $Pe$ increases beyond a critical value, and this critical value is consistent with that obtained through linear stability analysis. When the homogeneous steady state loses stability, a range of patterns emerge. Representative examples of these patterns are displayed in Fig.~\ref{attractors}. The patterns are characterized through kymographs, namely the heat map of the concentration of species $A$ in space and time (left panels in Fig.~\ref{attractors}). Additionally, the temporal evolution at typical locations in space is characterized by dynamical attractors in the phase space of the concentrations of the two chemical species (right panels in Fig.~\ref{attractors}). Typically, for smaller P\'eclet numbers oscillatory states arise, yielding regular limit cycle attractors in phase space. On the other hand, large P\'eclet numbers give rise to enhanced irregularity in the spatio-temporal patterns. For instance, in the representative examples in the figure, for $Pe=13$, the emergent space-time patterns are nearly pulsatory. On the other hand, for higher P\'eclet numbers such as $Pe=48$, one obtains a state characterized by long-term aperiodic temporal dynamics. This dynamics gives rise to a chaotic attractor in phase space, exhibiting a complex irregular geometric structure, in stark contrast to the regular simple limit cycle for lower P\'eclet numbers.

\begin{figure*}
\includegraphics[width=\textwidth]{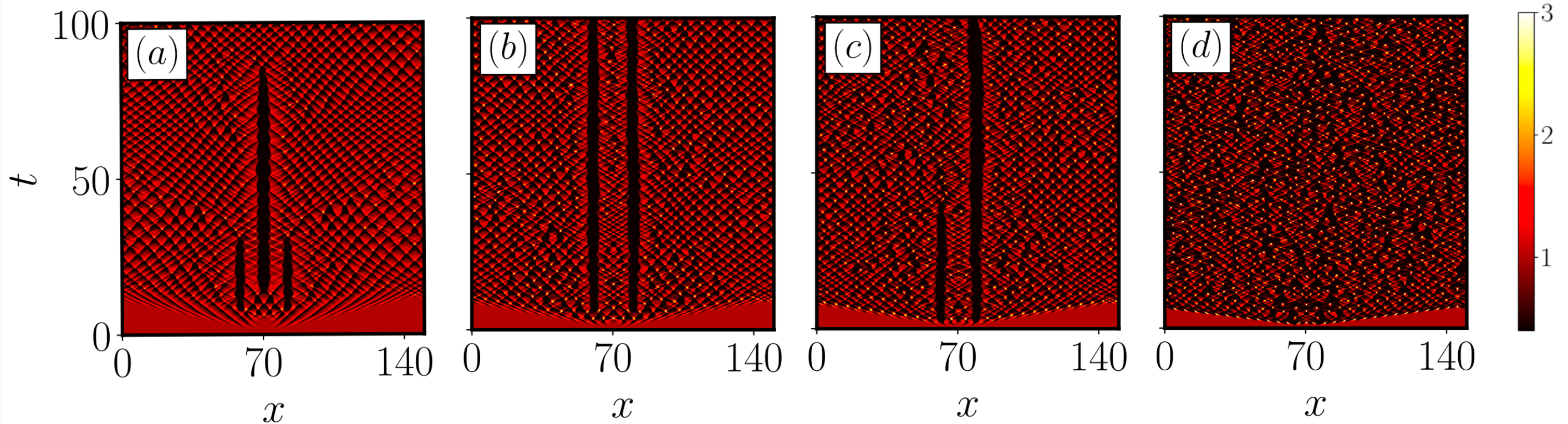}
  \caption{Emergence of an illustrative sequence of space-time patterns, ranging from transient Chimera states to irregular arrhythmic spatiotemporal patterns, in a system with $R=1$, $\alpha=0.1$, $\beta=3$, for increasing P\'eclet numbers: (a) $Pe = 27$, (b) $Pe=35$, (c) $Pe=39$,(d) $Pe=48$.
  }
    \label{pe_increasing}
\end{figure*}

\begin{figure}
    \centering
    \includegraphics[width=\textwidth]{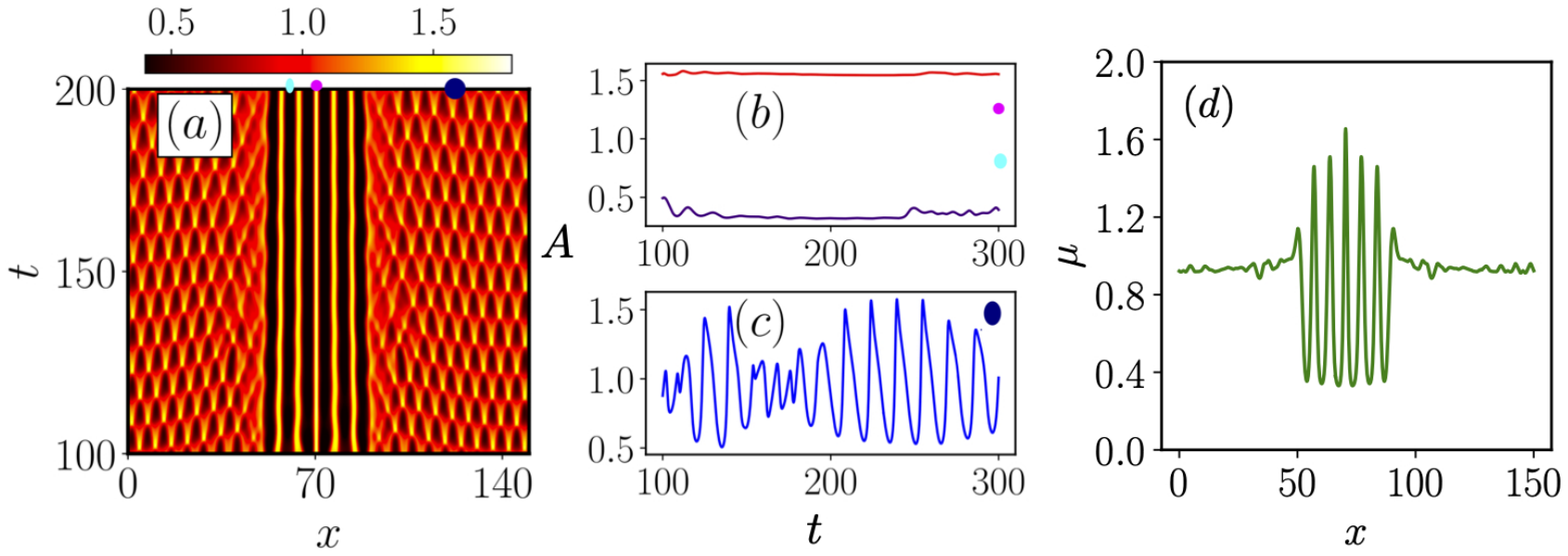}
    \caption{Long-lived two-regime transient chimera state. (a) Kymograph displaying the spatio-temporal pattern. (b) and (c) show the time evolution of concentration $A$, at locations in space marked by different symbols in the kymograph. (d) displays the mean value $\mu$ of the concentration of $A$ at different sites across the lattice, showing markedly different values in the distinct spatial domains, thereby serving as a good quantitative measure of the chimera state. The parameters here are: $\beta=1.2$, $\alpha=0.1$, $R=1$ and $Pe=13$.}
    \label{2regime}
\end{figure}

\begin{figure*}
    \centering
    \includegraphics[width=\textwidth]{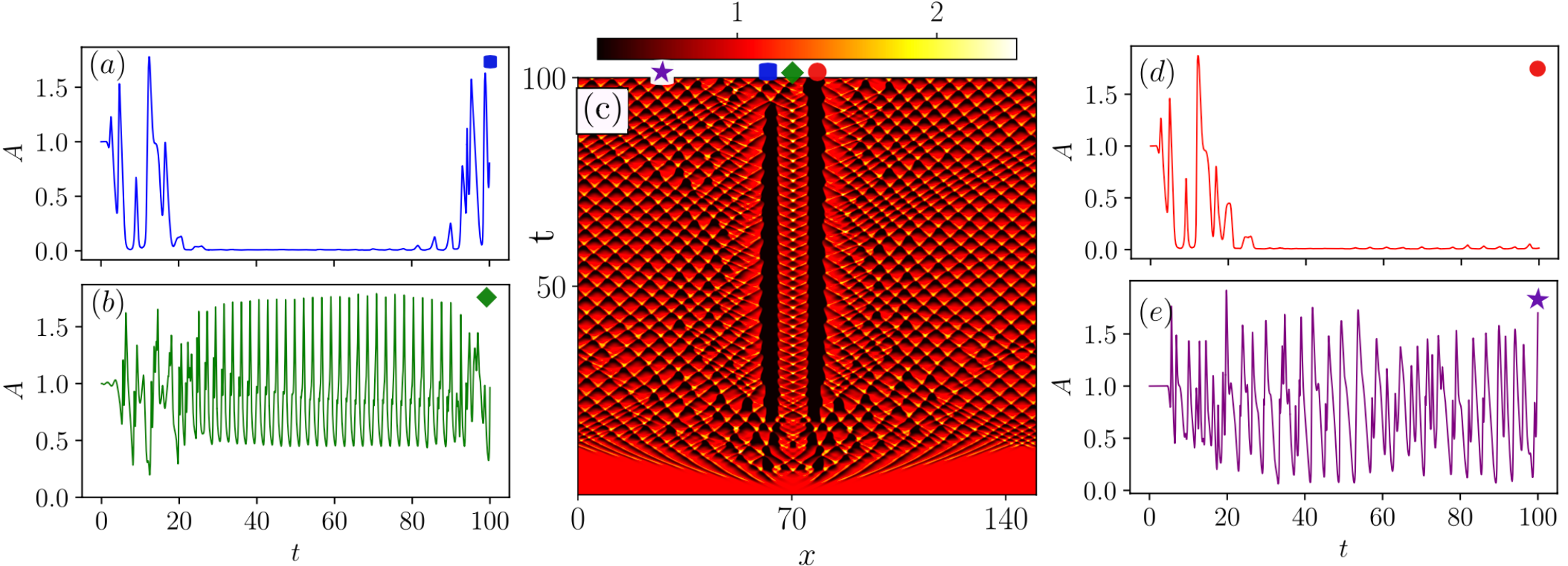}
   \caption{\raggedright Three-regime transient chimera state. The kymograph displays the spatiotemporal pattern formation from time 0 to 100. The panels on the sides display the time evolution of concentration $A$ at locations in space marked by different symbols in the kymograph.
 Here the parameters are: $\beta=3$, $\alpha=0.1$, $R=1$ and $Pe=32$. Note that the asymptotic state of the system is pulsatory, displaying regular oscillations in time and space.}
     \label{3_regime_chimera}
\end{figure*}

Figure~\ref{pe_increasing} shows a representative sequence of space-time patterns that emerge for increasing  P\'eclet numbers beyond the regime of regular pulsatory patterns. This illustrative set of patterns reveals the existence of interesting transient chimera states for moderately high P\'eclet numbers. 
Typically, the transient chimera states are non-stationary, and they appear, disappear and re-appear. This sequence repeats over long times before finally terminating and settling down to the asymptotic oscillatory pattern. In the context of transient chimeras, it is interesting to note that chimeras in the human cortex are often transient, and identifying them plays a role in predicting epileptic seizures \cite{lainscsek2019cortical}.

\begin{figure*}
\includegraphics[width=\textwidth]{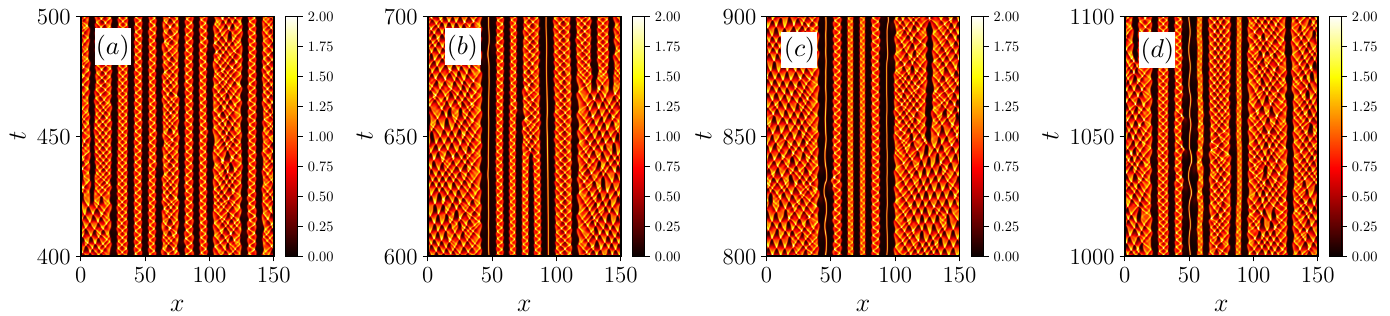}
    \caption{Kymographs displaying robust multi-head chimera states, for parameter values $Pe=21$, $\beta=2$, $R=1$, $\alpha=0.1$. Here time ranges from $400$ to $1100$, across the four panels,  from left to right.}
    \label{robust_chimera}
\end{figure*}

Now, a wide variety of chimera states emerge dependent on initial states, for large enough systems, with the persistence of transient chimeras increasing with increasing length-scale $L$ of the system. A representative example of a transient two-regime chimera state is shown in Fig.~\ref{2regime}. Here, two types of patterns coexist. We have homogeneous spatial patches where the concentration is nearly stationary for significant times, co-existing with alternating patches having a high and low concentration, i.e. nearly homogeneous stationary patches coexist with a spatial region exhibiting oscillatory space-time patterns. An example of a three-regime chimera state is displayed in Fig.~\ref{3_regime_chimera}. Here, nearly homogeneous steady patches coexist with two separate spatial regions exhibiting distinct oscillatory patterns. The time evolution of concentrations at representative locations in the distinct regimes is explicitly shown in Figs.~\ref{2regime}-\ref{3_regime_chimera}. Notice that in the multi-domain chimera displayed in Fig.~\ref{3_regime_chimera}, the dynamics in the domains (b) and (e), while qualitatively similar, exhibit different degrees of regularity and frequencies of the oscillatory behaviour which can be discerned from the dominant frequencies in the Fourier spectra and the level of the noisy background vis-a-vis the peaks. Further, notice that the oscillatory spatial domains are separated by domains of temporal stationarity (evident as dark vertical domains in the kymographs), and these domains are also of significantly different spatial extents. Interestingly, in the coherent regimes of the chimaera states, the net flux and the magnitude of the growth term are close to zero, and so these regimes behave like a spatial barrier, with soliton-like waves moving, colliding and passing through each other, but reflected back at the boundary of the coherent regimes. So these waves are trapped within a domain between the two coherent regimes, forming a distinct spatial domain with temporal behaviour very different from the coherent domain, i.e. the chimera states are characterized by the coexistence of irregular motion of soliton-like waves and regions of coherence with almost zero concentration. Additionally, we find the emergence of robust chimera states, as shown in Fig.~\ref{robust_chimera}.

\begin{figure}
    \centering
 \includegraphics[width=0.5\linewidth]{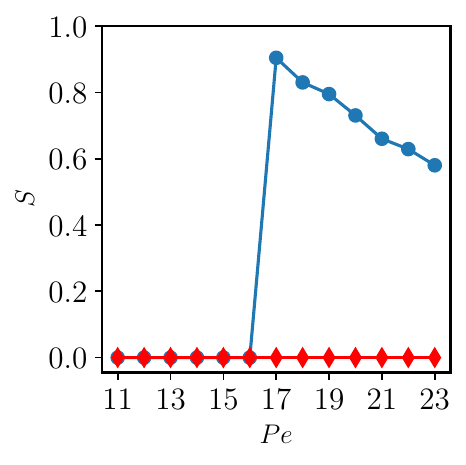}
    \caption{Dependence of the strength of incoherence $S$ (see text for definition) on the P\'eclet number. Here $\beta=2$, $\alpha=0.1$, and $R=1$ (blue circles) and $R=0.25$ (red diamonds). The quantity $S$ takes values ranging from $0$ (characteristic of coherent states) to $1$ (characteristic of completely incoherent states), and takes values within the range $(0,1)$ for chimera states. The emergence of chimera states after $Pe \sim 15$ is clearly evident for $R=1$, while it does not occur for $R=0.25$.}
    \label{incoherence}
\end{figure}

In parameter regimes beyond those supporting chimera states, almost all initial perturbations give rise to irregular arrhythmic oscillatory spatio-temporal patterns, illustrative examples shown in Figs.~\ref{attractors} and Fig.~\ref{pe_increasing}d.  So when the P\'eclet number is very small, we obtain stable homogeneous steady states. As $Pe$ increases, chimera states emerge and are found over a considerable range of P\'eclet numbers. For very high $Pe$, chimera states again disappear, and irregular arrhythmic patterns, akin to spatiotemporal chaos, are commonly observed. So, one can conclude that a moderately high P\'eclet number is most conducive to chimera-like states.

In the phase diagrams obtained through stability analysis (cf. Figs.~\ref{Pe_R}-\ref{Pe_beta}), the regime of oscillatory instability is most conducive to chimera states. Notice that this regime is most extensive for large $R$, implying that the possibility of observing chimeras is enhanced by larger nonlinearity. This connection with the phase diagram is also consistent with the observation of chimeras at moderate P\'eclet numbers but not at high $Pe$, where the oscillatory instability gives way to stationary instability. Lastly, the asymmetry parameter $\beta$ needs to be larger than $1$ to obtain chimeras. This observation is again linked to the fact that the oscillatory instability exists only for $\beta >1$.

\begin{figure}
    \centering
    \includegraphics[width=\textwidth]{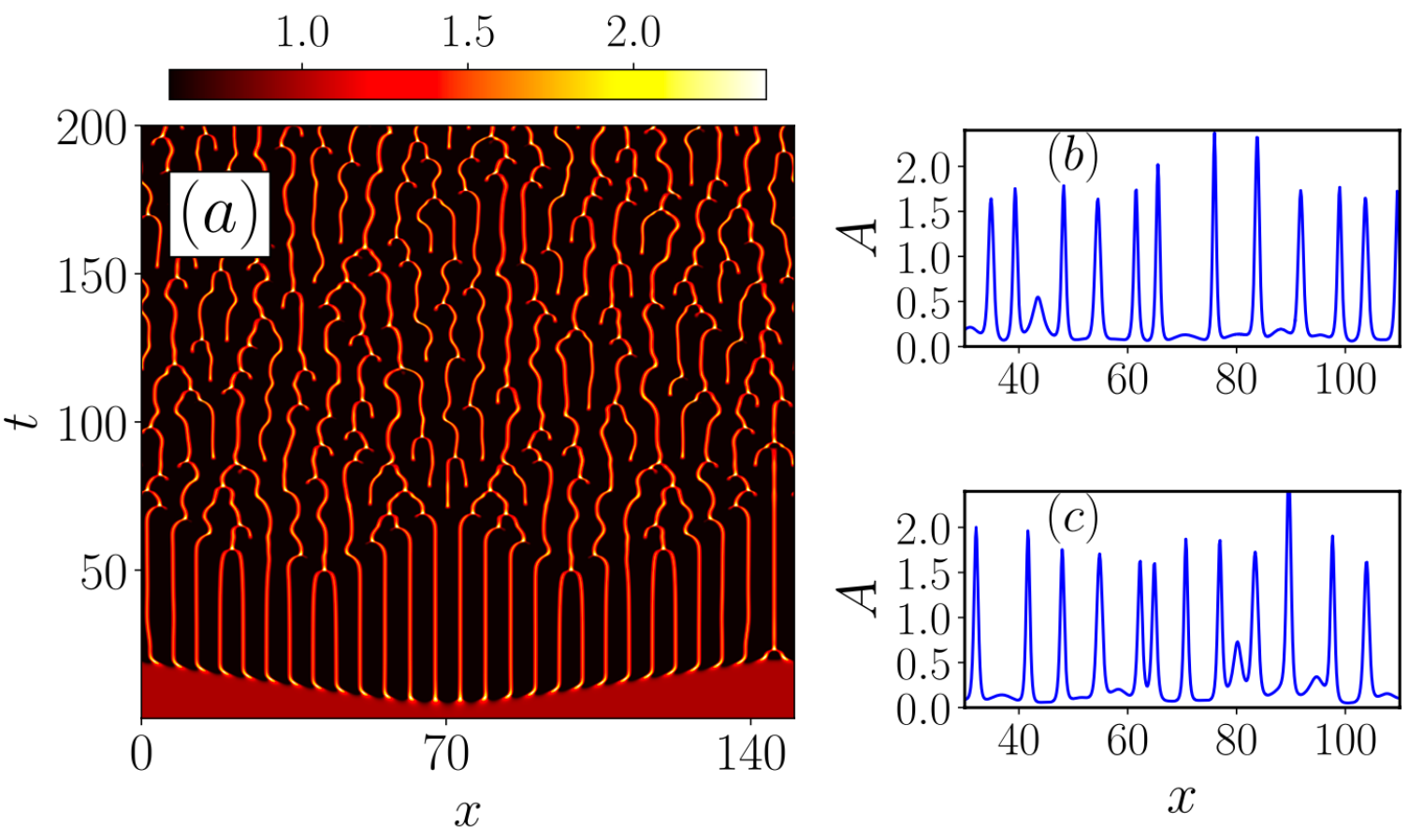}
    \caption{Merging-emerging spatiotemporal dynamics. (a) Kymograph showing merging-emerging dynamics, for parameters $Pe=20$, $R=1$, $\beta=1.2$ and $\alpha=0.1$. Panels (b) and (c) show the spatial variation of concentration $A$ at two instants of time $t=190$ and $t=194$. The emergence and annihilation of concentration peaks are clearly visible in the spatial profile.}
    \label{merge}
\end{figure}

To further quantify the prevalence of chimeras, we have computed the strength of incoherence which serves as a good order parameter to capture the existence of chimeras and provides a measure to characterize chimera states in complex systems. This quantity is defined as \cite{SI}:
$S = 1 - \frac{\sum_x H(\bar{A})}{N}, H(\bar{A}) = \Theta[\delta - \bar{A}]$, where $\bar{A} = \langle A(x) \rangle - \langle A_c \rangle$. Here, $\bar{A}$ measures the deviation of species densities at $x$ from the coherent subpopulation, $\langle A_c\rangle$ represents the average amplitude of the coherent subpopulation, and $\delta$ is a threshold value. The quantity $S$ takes values ranging from $0$ (characteristic of coherent states) to $1$ (characteristic of completely incoherent states), and takes values within the range $(0,1)$ for chimera states, indicating the coexistence or coherence and incoherence. So $S$ can serve as an ``order parameter'' indicating the occurrence of chimeras. Fig.~\ref{incoherence} shows this quantity over a range of P\'eclet numbers. Clearly, $S=0$ when $R$ is low (specifically $R=0.25$ in the figure). On the other hand, chimeras occur in a window of P\'eclet number when $R$ is sufficiently large (specifically $R=1$), where we have a homogeneous steady state till $Pe \sim 11$ (with $S=0$), a regular pulsatory state for $Pe \sim 12-15$ (with $S=0$), and chimeras beyond $Pe \sim 15$ (with $0 < S < 1$).

Finally, we observe merging-emerging solitonic structures in the space-time evolution of the concentrations. Such patterns involve the spontaneous emergence of localized concentration peaks (referred to as ``emerging") and the coalescing of two existing localized concentration peaks (referred to as ``merging"). This class of patterns is illustrated in Fig.~\ref{merge}, which shows the destabilization of the initial near-uniform spatial profile into multi-peak aggregations, which evolve through a sequence of emerging and merging events. Such patterns are reminiscent of solitonic defects, where the soliton-like structures emerge, persist for some time and then may disappear, merge with other neighbouring solitonic defects or split into solitonic sub-structures as the system evolves. These patterns are most predominant in the regime of stationary instability. So low asymmetry $\beta$ and high P\'eclet numbers are most likely to yield them. Such merging and emerging space-time structures are similar to patterns observed in the dynamics of a model for chemotaxis incorporating a logistic cell growth term \cite{painter2011spatio}. Further, such patterns have also been observed in a model system that mimics nonlinear cell-diffusion \cite{wang2007classical}.

In this study, we considered the widely used periodic boundary conditions. Prototypically, periodic boundary conditions are chosen to approximate large systems and eliminate edge effects in simulations, and for these reasons, they have been broadly employed in the study of model systems mimicking dynamics within a fluid environment or membrane-like structure. Of course, the influence of different boundary conditions on emergent patterns is an open problem of considerable relevance and holds much potential for future investigations. 

\section{Two-dimensional Systems}

We have extended our simulations to two dimensions. The governing equations are now expressed as
\begin{equation}
 \partial_t A = - \nabla \cdot (\mathbf{\mathbf{v}}A) + D\nabla^2 A + rA(1 - \frac{A}{K})\\ 
 \end{equation}
 \begin{equation}
\partial_t I = - \nabla \cdot( \mathbf{\mathbf{v}}I) + \alpha D\nabla^2 I\\
\end{equation}
\begin{equation}
    \nabla \cdot \mathbf{\sigma} = \gamma\mathbf{v}
\end{equation}
\begin{equation}
    \sigma = \sigma_p + \sigma_a\mathbb{I}
\end{equation}
\begin{equation}
    \sigma_p = [ \nabla \mathbf{v} + (\nabla \mathbf{v})^T - \frac{2}{d}(\nabla \cdot \mathbf{v})\mathbb{I}] + \eta_v(\nabla \cdot \mathbf{v})\mathbb{I}
\end{equation}
\begin{equation}
    \sigma_a = \sigma_0 f(A,I)
\end{equation}
Here, $\eta$ and $\eta_v$ represent the shear and bulk viscosity, $d$ is the dimension (here, $d=2$). Specifically, we choose $\eta_v$ = $3\eta$~\cite{jenkins1997effective,staddon2022pulsatile} and use a semi-implicit spectral scheme to simulate the dynamics given by the equations above. Pattern formation is explored mostly in the square domain, with periodic boundary conditions, under a small local perturbation. In the dimensionless form, the equations are given below: 
\begin{equation}
 \partial_t A = - \nabla \cdot (\mathbf{\mathbf{v}}A) + \nabla^2 A + RA(1 - A)\\ 
 \end{equation}
 \begin{equation}
\partial_t I = - \nabla \cdot( \mathbf{\mathbf{v}}I) + \alpha \nabla^2 I\\
\end{equation}
 \begin{equation}
\kappa\partial^2_x u + \epsilon\frac{\partial^2v}{\partial x\partial y} +Pe\frac{\partial f}{\partial x} + \partial^2_y u   = u\\
\end{equation}
 \begin{equation}
\kappa\partial^2_y v + \epsilon\frac{\partial^2u}{\partial x\partial y} +Pe\frac{\partial f}{\partial y} + \partial^2_x v   = v\\
\end{equation}
where, $\kappa = 1 + \epsilon$ with  $\epsilon = \frac{\eta_v}{\eta}$. Also, $\mathbf{v}=(u,v)$.

\begin{figure*}
\includegraphics[width=\textwidth]{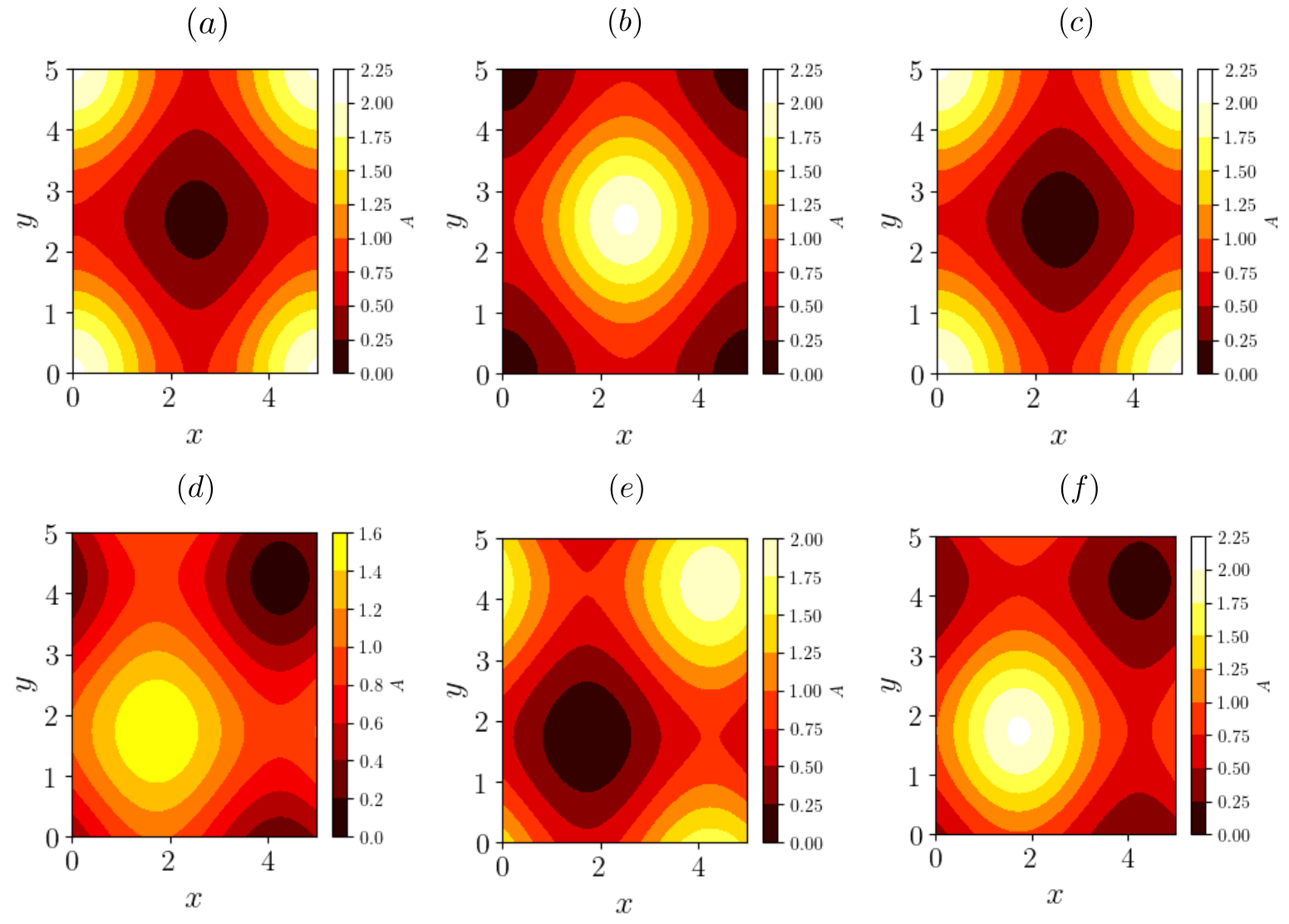}
    \caption{Snapshots of the $2$-dimensional domain, at different instants of time: (top row) Pulsatory spatiotemporal patterns under perturbation at one central site. (Bottom row) Two distinct pulsatory spatial domains, with alternating high and low concentrations, under perturbation at two random sites. Here the heat map show the concentration of $A$, and parameter values are $Pe=40$, $\beta=3$ , $R=1$ and $\alpha=0.1$. The step-size $\Delta t$ = $0.001$ and $\Delta x$ = $0.03$ in the simulation, with $L_x=L_y=5$. 
   }
\label{2d_pulsatory}
\end{figure*}

\begin{figure*}
\includegraphics[width=\textwidth]{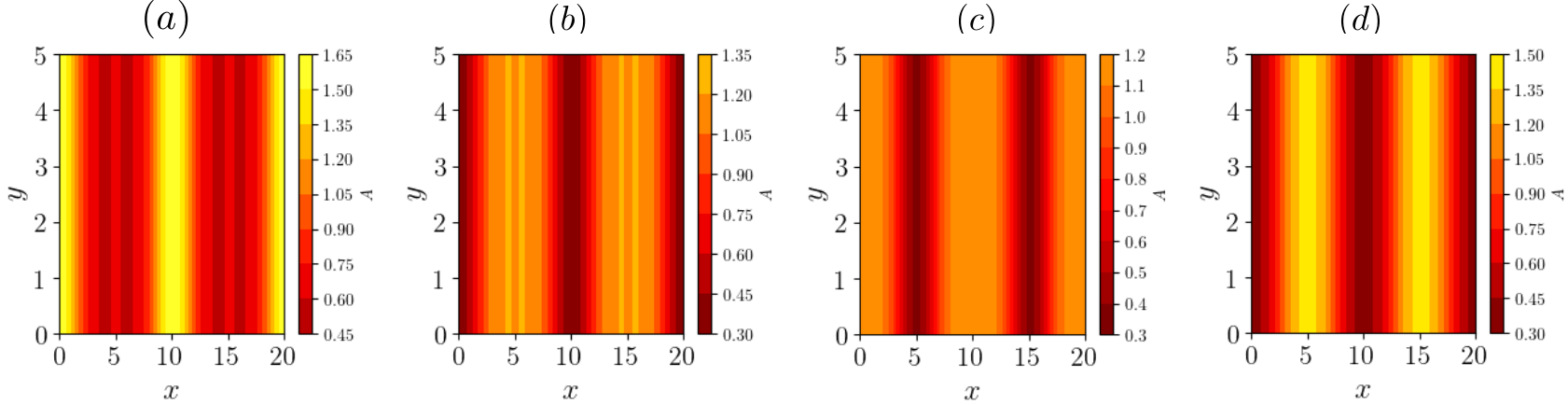}
    \caption{Snapshots of a $2$-dimensional rectangular domain, with $L_x=20, L_y=5$, at different instants of time, showing pulsatory band-like spatiotemporal patterns. Here the system displays synchronized vertical domains, and splits into multi-clusters along the horizontal axis.  
    }
    \label{band}
\end{figure*}

\begin{figure*}
\includegraphics[width=\textwidth]{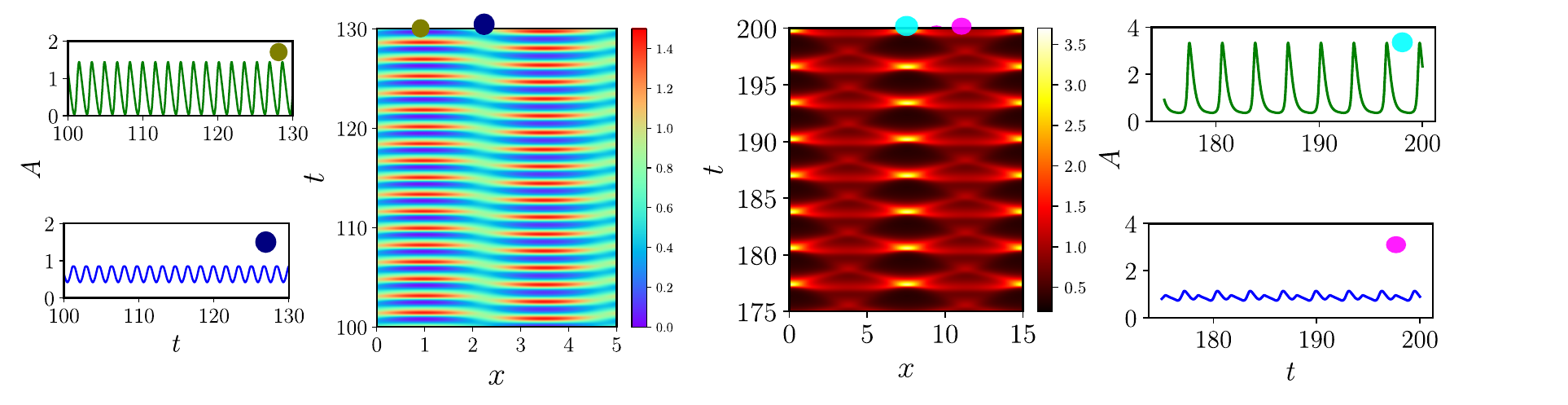}
    \caption{Chimera state in two-dimensional systems: the kymograph shows a horizontal section in the square lattice for  $Pe=80$, $\beta=3$, $R=1$, $\alpha=0.1$, $L_x=L_y=5$ 
    (Left) and $Pe=50$, $\beta=3$, $R=1$, $\alpha=0.1$, $L_x=L_y=15$ (Right),  alongside time series at two representative spatial locations placed in the two distinct domains, as indicated by different symbols on the kymograph.  
    }
    \label{2d_chimera_kymo}
    \end{figure*}

As in the $1$-dimensional case detailed earlier, we observe homogeneous steady states and regular pulsatory patterns, as well as stationary inhomogeneous patterns and symmetry-breaking amplitude chimera states (Figs.~\ref{2d_pulsatory}-\ref{2d_chimera_kymo}). In particular, we present a few illustrative spatio-temporal patterns in Fig.~\ref{2d_pulsatory}, showing the emergence of diverse pulsatory patterns. For instance, for a square domain with a perturbation at a random site, we obtain periodic circular patterns with a circle that alternate between high and low concentrations, evident as periodically changing bright and dark circles in the heat map of the domain. If perturbations are applied at two random sites, alternating bright-dark circular structures emerge. We have also explored pattern formation in a rectangular domain. We find interesting pulsatory band-like structures (see Fig.~\ref{band}), reminiscent of splay patterns \cite{xie2015twisted}, as well as chimera states with two regimes with distinct amplitudes (see Fig.~\ref{2d_chimera_kymo}). {Note that we have explored system sizes an order of magnitude larger than those displayed in these figures, and we find that qualitatively similar patterns emerge for larger systems as well.}

So it is apparent that two-dimensional active fluid systems offer a rich repertoire of spatiotemporal patterns, ranging from homogeneous steady states to chimera states. Our results here can then trigger future avenues of exploration of chimeras in active fluid systems in domains of different geometries under different boundary conditions and perturbations.

\section{Conclusions}

We explored a two-component system, where the chemical species interact with the active fluid medium via active stress gradients, and the growth of the fast-diffusing species is modelled by a nonlinear logistic term. We investigated the formation of space-time patterns, both analytically and through extensive numerical simulations. Our key results are summarized below.

First, increasing the P\'eclet number beyond a critical value leads to the loss of stability of the homogeneous steady state, i.e. instability is induced when the rate of advection is sufficiently greater than the rate of gradient-driven diffusion. Further, the asymmetry of the activator and inhibitor species in our two-component system, as reflected by the parameter $\beta$, also has a crucial effect on the emergent dynamical phases. Increasing asymmetry yields three distinct dynamical phases, while low asymmetry does not yield oscillatory instability. Lastly, most interestingly, we demonstrate that nonlinear growth aids the stability of the homogeneous steady state. This counter-intuitive phenomenon is evident from the enlarged area of stability of the homogeneous steady state in the phase diagrams for larger-scaled nonlinear growth parameter $R$. Specifically, we also show that the critical P\'eclet number for the onset of instability rises approximately linearly with $R$, leading to an expansion of the homogeneous steady state region in parameter space. Our analytical results explicitly show the emergence of different dynamic phases, which are entirely corroborated by numerical simulations.

The second set of significant results arises in the region of parameter space where the homogeneous steady state loses stability. This region is explored extensively numerically, and complex eigenvalues for certain modes in the perturbed system are determined, indicating oscillatory instabilities. We find the emergence of diverse classes of patterns in this regime. These range from regular oscillatory space-time patterns to merging-emerging multi-peak aggregations reminiscent of solitonic defect-like structures and irregular space-time evolution characterized by chaotic attractors. Most interestingly, we find the emergence of chimera states, both long-lived transient and robust, and these chimeras are predominant in the region of oscillatory instability. Now, the dispersion curves for different modes of the perturbed system obtained through our linear stability analysis show that the region of oscillatory instability is enlarged in the presence of nonlinear growth. This implies the following important result: nonlinear growth enhances the probability of observing chimera states. These results can potentially aid experimentalists, as they can focus their search for chimera states on classes of systems more prone to such patterns.

\bigskip

\noindent
{\bf Appendix: Reduced-Order Analysis - Galerkin Projection and SVD}

\bigskip

{To complement the direct numerical simulations and linear stability analysis, we perform a reduced-order study using two approaches: (i) Galerkin mode truncation and (ii) Singular Value Decomposition (SVD). While the Galerkin method provides a low-dimensional approximation of the governing PDEs by projecting them onto a truncated set of orthogonal modes, SVD offers a data-driven decomposition of the spatio-temporal patterns, revealing the hierarchy of dominant modes.}

\begin{figure}
\begin{center}
\includegraphics[width=\textwidth]{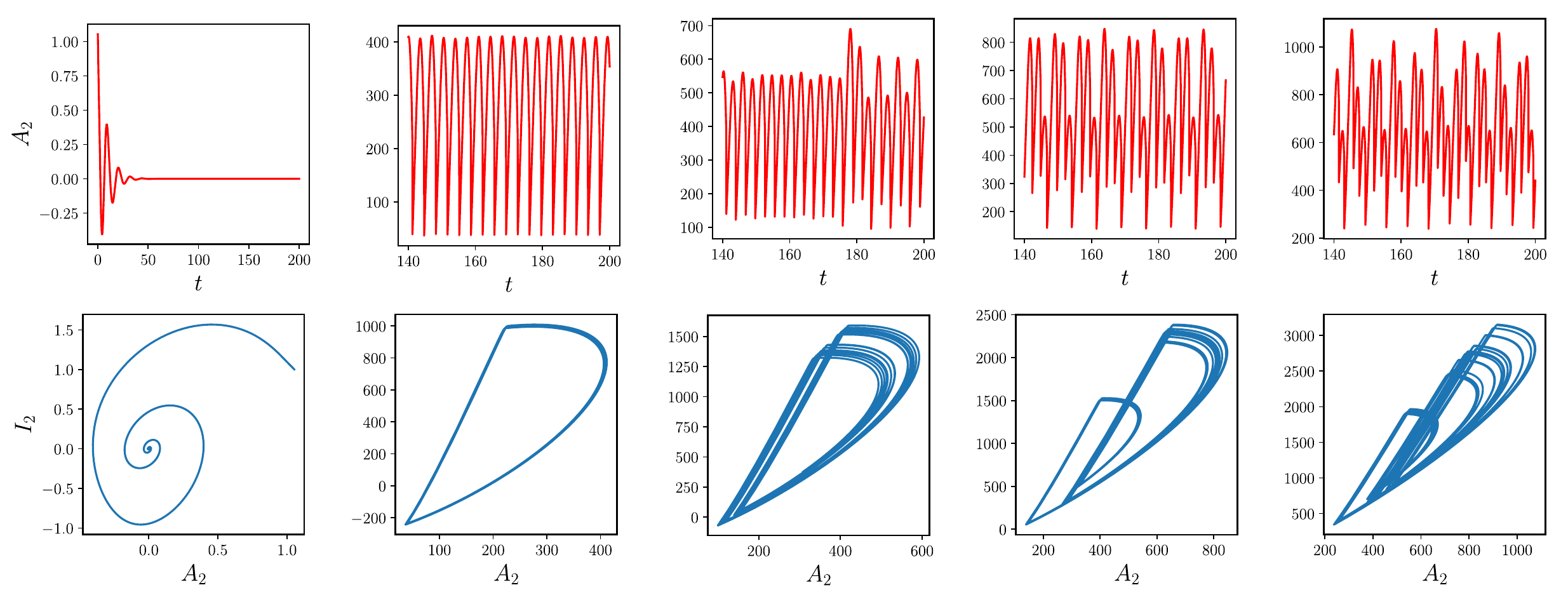}
\end{center}
\caption{\raggedright {$2$-mode coupling results: upper row displays the time series of $A_2$, and the lower row displays the phase portraits of the temporal coefficients in the space of $A_2-I_2$, for the parameters are (left to right): $Pe=20, 40,50,58,68$, with $\beta=1.75$, $R=1$, $\alpha=0.1$, $L=50$.}}
    \label{merge11}
\end{figure}

\begin{figure}
    \centering
    \includegraphics[width=\textwidth]{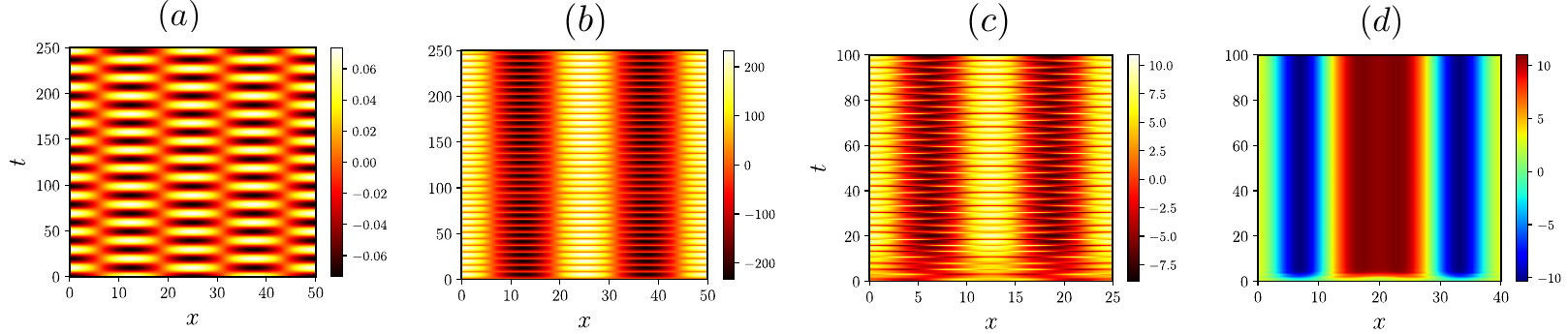}
    \caption{{(a) Kymograph of Pulsatory patterns of the perturbation ($A_p$)from $Pe=27,\beta=3,R=1,\alpha=0.1,L=50$ and (b) image represents the amplitude Chimera states of the perturbation($A_p$) for $A$ with $Pe=33,\beta=3,R=1,\alpha=0.1,L=50$ (obtained from Mode 2 coupling). (c) Kymograph of $A$ for $Pe=11,\beta=2,R=1,\alpha=0.1,L=25$ (Chimera states) and (d) multi-valued Stationary states kymograph for $A$ for $Pe=12,\beta=2,R=1,L=40$ (obtained from $3$ mode truncation analysis).}}
    \label{merge0}
\end{figure}

\bigskip

\noindent
{\bf A. Galerkin Mode Truncation Analysis}

\bigskip

{We first apply Galerkin projection~\cite{fletcher1984computational} to approximate the temporal evolution of the dominant modes. The concentration fields are expanded around the homogeneous steady state:
\begin{equation}
    A = A_0 + A_p(x,t), \quad I = I_0 + I_p(x,t), \quad v = v_0 + v_p(x,t).
\end{equation}
Cosine functions are chosen as basis functions due to their orthogonality under periodic boundary conditions. Considering the first two modes, the perturbations are written as:
\begin{equation}
    \epsilon(x,t) = \epsilon_1(t)\cos\left(\frac{2\pi x}{L}\right) + \epsilon_2(t)\cos\left(\frac{4\pi x}{L}\right),
\end{equation}
\begin{equation}
    A_p(x,t) = A_1(t)\cos\left(\frac{2\pi x}{L}\right) + A_2(t)\cos\left(\frac{4\pi x}{L}\right),
\end{equation}
\begin{equation}
    I_p(x,t) = I_1(t)\cos\left(\frac{2\pi x}{L}\right) + I_2(t)\cos\left(\frac{4\pi x}{L}\right),
\end{equation}
where \( \epsilon = \partial v / \partial x \) denotes the strain rate. For simplicity, we use a linearized active stress function:
\begin{equation}
    f(A,I) = (1 + \beta)\frac{A}{A_S} + (1 - \beta)\frac{I}{I_S}.
\end{equation}}


{Using orthogonality, the evolution equations for the amplitudes \( A_i, I_i \) are derived as:
\begin{equation}
    \epsilon_1 = \frac{-Pe \, k^2 \left[(1+\beta)\frac{A_1}{A_S} + (1-\beta)\frac{I_1}{I_S}\right]}{1 + k^2},
\end{equation}
\begin{equation}
    \epsilon_2 = \frac{-4Pe \, k^2 \left[(1+\beta)\frac{A_2}{A_S} + (1-\beta)\frac{I_2}{I_S}\right]}{1 + 4k^2},
\end{equation}
\begin{equation}
    \frac{dA_1}{dt} = -\epsilon_1 + \frac{3A_1\epsilon_2}{4} + \frac{3A_2\epsilon_1}{2} - k^2 A_1 - R A_1 - R A_1 A_2,
\end{equation}
\begin{equation}
    \frac{dA_2}{dt} = -\epsilon_2 - A_1 \epsilon_1 - k^2 A_2 - R A_2 - \frac{R A_1^2}{2},
\end{equation}
\begin{equation}
    \frac{dI_1}{dt} = -\epsilon_1 + \frac{3I_1\epsilon_2}{4} + \frac{3I_2\epsilon_1}{2} - \alpha k^2 I_1,
\end{equation}
\begin{equation}
    \frac{dI_2}{dt} = -\epsilon_2 - I_1 \epsilon_1 - \alpha k^2 I_2.
\end{equation}
}

{This two-mode truncation captures key qualitative features of the system, including steady states, oscillations, and even more interestingly, amplitude chimera states (see Figs.~\ref{merge11}-\ref{merge0}). However, it is limited in scope and fails to preserve certain intrinsic properties of the full PDEs, such as global conservation and boundedness~\cite{schein2021preserving}. For instance, unphysical negative concentration values may arise in the reduced model. Extending the analysis to three modes (yielding six coupled ODEs) mitigates such artifacts and provides better agreement with the full numerical solutions. For example, chimera states are observed for \( Pe = 11, \beta = 2, R = 1, \alpha = 0.1, L = 25 \) in the three-mode model (see Fig.~\ref{merge0}), consistent with direct PDE simulations where chimera states emerge at \( Pe = 17 \). Phase portraits of the amplitude variables \( (A_1, A_2, A_3, I_1, I_2, I_3) \) exhibit a mix of regular (limit-cycle-like) and complex (folding and stretching) structures, indicating quasi-periodic or aperiodic behavior and hinting at the presence of chimera-like states (see bottom row of Fig.~\ref{merge11}).
}

\begin{figure}
    \centering
    \includegraphics[width=\textwidth]{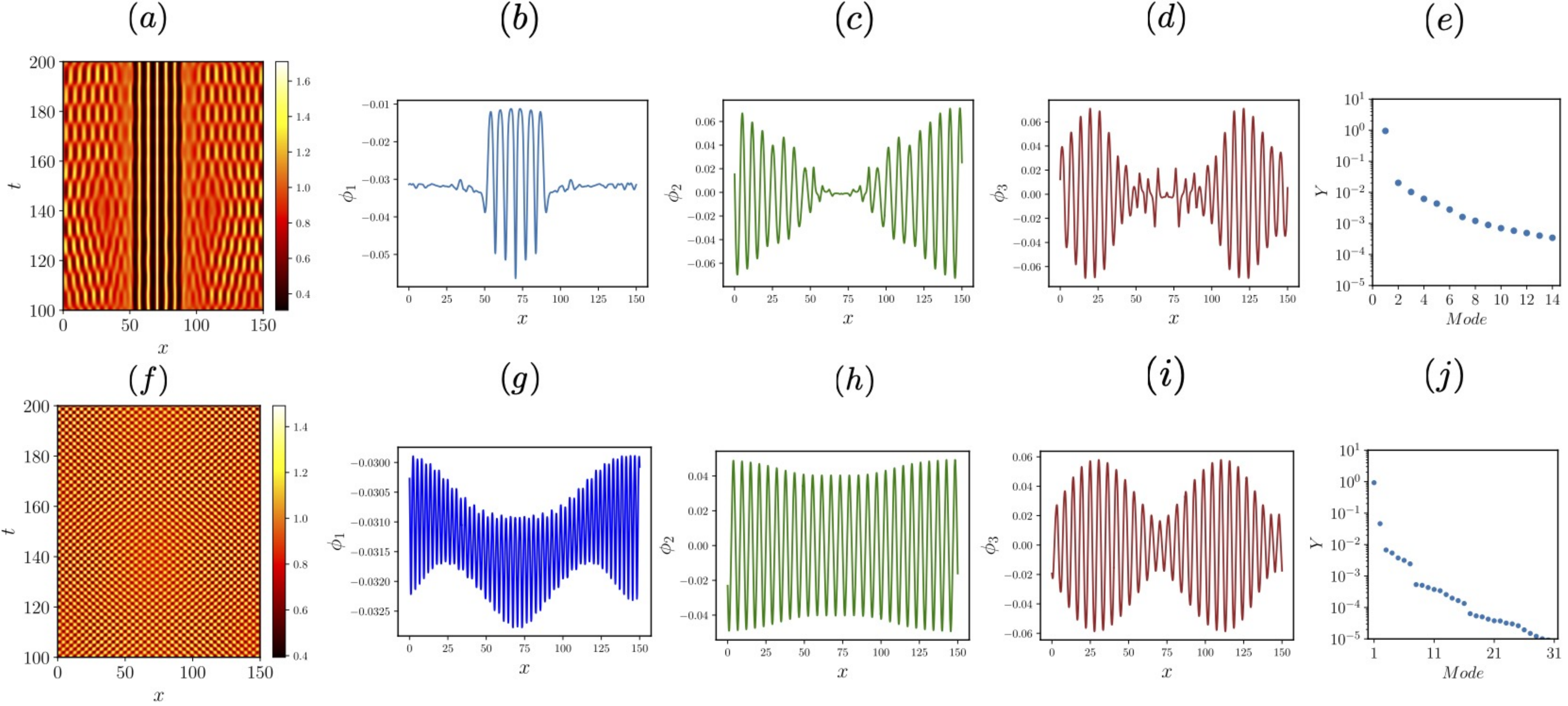}
    \caption{{(a-e) Reconstructed kymograph with first $3$ dominant modes obtained from SVD, three optimal basis functions and relative strength of each basis modes for $Pe=13,\beta=1.2,R=1,\alpha=0.1$. (f-j) Reconstructed kymograph with first $3$ dominant modes obtained from SVD, three optimal basis functions and relative strength of each basis modes for $Pe=15,\beta=3,R=1,\alpha=0.1$.}}
    \label{merge1}
\end{figure}

\begin{figure}
    \centering
    \includegraphics[width=\textwidth]{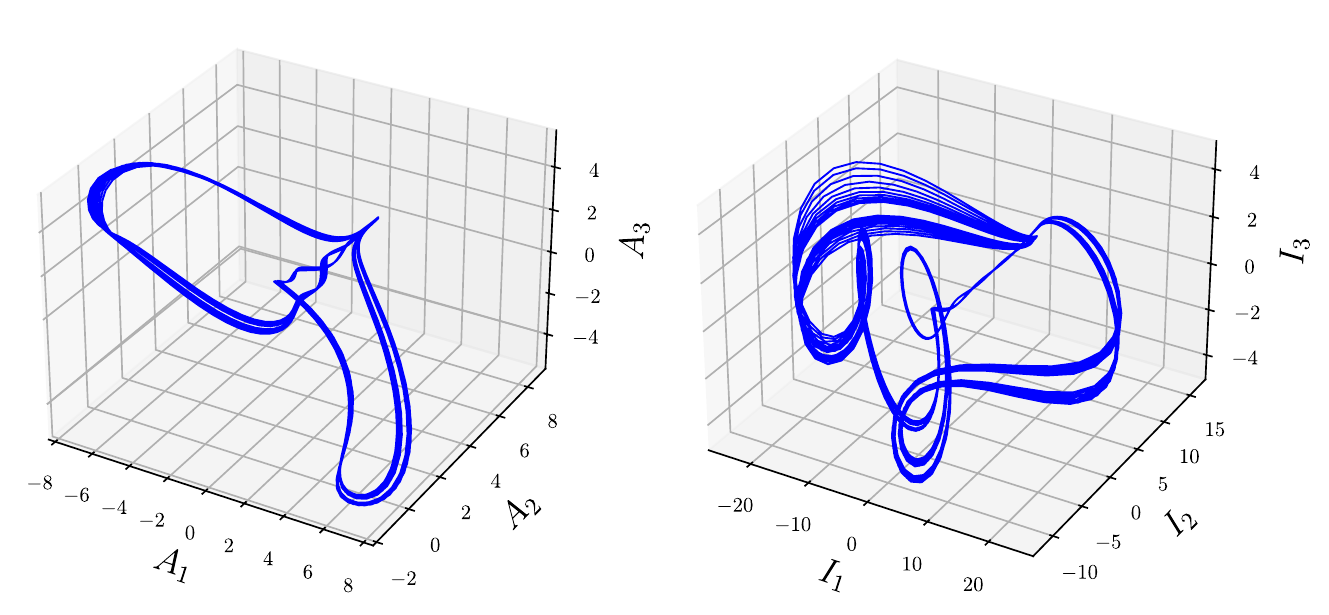}
    \caption{{3D attractors of $A_1,A_2,A_3$ and $I_1,I_2,I_3$ corresponding to 3 mode coupling, for $Pe=11,\beta=2,R=1,\alpha=0.1,L=25$.}}
    \label{merge28}
\end{figure}
\bigskip

\noindent
{\bf B. Singular Value Decomposition (SVD)}

\bigskip

{While the Galerkin approach imposes a predefined basis, the SVD provides a data-driven decomposition of the spatio-temporal fields. We apply SVD~\cite{shlizerman2012proper} to the simulation data to identify the optimal orthogonal modes and their relative strengths. For a data matrix \( A(x,t) \), the SVD is expressed as:
\begin{equation}
    A(x,t) = \sum_{j=1}^{N} \sigma_j \, \phi_j(x) \, \psi_j(t),
\end{equation}
where \( \phi_j(x) \) and \( \psi_j(t) \) are spatial and temporal modes, respectively, and \( \sigma_j \) are singular values. The normalized energy contribution of mode \( j \) is:
\begin{equation}
    Y_j = \frac{\sigma_j^2}{\sum_{i=1}^N \sigma_i^2}.
\end{equation}
}

{The SVD analysis reveals that in regimes of simple pulsatory dynamics (e.g., \( Pe = 15, \beta = 3, R = 1, \alpha = 0.1 \)), the first mode dominates (\( Y_1 \approx 1 \)) while higher modes are negligible. In contrast, for chimera states (e.g., \( Pe = 13, \beta = 1.2, R = 1, \alpha = 0.1 \)), the first three modes capture over \( 97\% \) of the energy, with \( \phi_1 \) representing coherent spatial regions and \( \phi_2, \phi_3 \) capturing oscillatory variations. For more complex regimes such as spatio-temporal chaos or merging-emerging patterns (e.g., \( Pe = 18, \beta = 1.2, R = 1, \alpha = 0.1 \)), up to \( \sim 200 \) modes may be required for accurate reconstruction. Here, the energy spectrum \( Y_j \) decays slowly, indicating strong contributions from higher-order modes (see Fig.~\ref{merge1}).
}

{Examining the temporal coefficients \( (A_1, A_2, A_3) \) and their phase portraits provides additional insights: pulsatory states display regular closed loops (limit cycles), while chimera and chaotic states show folding and stretching, reflecting the coexistence of coherent and incoherent dynamics (see Fig.~\ref{merge28}). Thus, SVD complements the Galerkin analysis by providing a compact, data-driven framework to dissect the hierarchical structure of spatio-temporal patterns.
}

\bigskip

\section*{References}

\bibliography{Chimera}
\bibliographystyle{iopart-num}

\end{document}